# Comparative Study on Agile software development methodologies

A B M Moniruzzaman
abm.mzkhan@gmail.com

Dr Syed Akhter Hossain
aktarhossain@daffodilvarsity.edu.bd

**Abstract**

Today's business environment is very much dynamic, and organisations are constantly changing their software requirements to adjust with new environment. They also demand for fast delivery of software products as well as for accepting changing requirements. In this aspect, traditional plan-driven developments fail to meet up these requirements. Though traditional software development methodologies, such as life cycle-based structured and object oriented approaches, continue to dominate the systems development few decades and much research has done in traditional methodologies, Agile software development brings its own set of novel challenges that must be addressed to satisfy the customer through early and continuous delivery of the valuable software. It's a set of software development methods based on iterative and incremental development process, where requirements and development evolve through collaboration between self-organizing, cross-functional teams that allows rapid delivery of high quality software to meet customer needs and also accommodate changes in the requirements. In this paper, we significantly indentify and describe the major factors, that Agile development approach improves software development process to meet the rapid changing business environments. We also provide a brief comparison of agile development methodologies with traditional systems development methodologies, and discuss current state of adopting agile methodologies. We speculate that from the need to satisfy the customer through early and continuous delivery of the valuable software, Agile software development is emerged as an alternative to traditional plan-based software development methods. The purpose of this paper, is to provide an in-depth understanding, the major benefits of agile development approach to software development industry, as well as provide a comparison study report of ASDM over TSDM.

*Author*

# 1. Introduction

A lot of people have been asking the question "What is Agile Software Development?" and invariably they get a different definition depending on who they ask. Here's a definition that conforms to the values and principles of the Agile Manifesto[1]. *An iterative and incremental (evolutionary) approach to software development which is performed in a highly collaborative manner by self-organizing teams within an effective governance framework with "just enough" ceremony that produces high quality solutions in a cost effective and timely manner which meets the changing needs of its stakeholders [6].* Agile software development is actually a group of software development methods based on iterative and incremental development, where requirements and solutions evolve through collaboration between self-organizing, cross-functional teams [4]. In 2001, the ''agile manifesto'' was written by the practitioners reveals which items are considered valuable by ASDMs [1]. As shown in Table 1.

| **More Valuable Items** | | **Less Valuable Items** |
|---|---|---|
| Individuals and Interactions | | Processes and tools |
| Working software | over | Comprehensive Documentation |
| Customer collaboration | | Contract negotiation |
| Responding to change | | Following a plan |

Table 1: Agile Manifesto (source: [1])

## 1.2 Research Review

Agile software development (ASD) is major paradigm, in field of software engineering which has been widely adopted by the industry, and much research, publications have conducted on agile development methodologies over the past decade. The traditional way to develop software methodologies follow the generic engineering paradigm of requirements, design, build, and maintain. These methodologies are also called waterfall–based taking from the classical software development paradigm. They are also known by many other names like plan–driven, (Boehm and Turner, 2004), [39]; documentation driven, heavyweight methodologies, and big design upfront, (Boehm, 2002), [16]. Boehm and Phillip [72] report that during their project development experience, requirements often changed by 25% or more. Due to constant changes in the technology and business environments, it is a challenge for TSDMs to create a complete set of requirements up front [26]. Williams and Cockburn, [18] also mentioned that one of problems of TSDMs is the inability to respond to change that often determines the success or failure of a software product.

The agile approach to software development is based on the understanding that software requirements are dynamic, where they are driven by market forces (Fowler, 2002; Cockburn & Highsmith, 2001); [16], [36]. Agile systems development methods emerged as a response to the inability of previous plan-driven approaches to handle rapidly changing environments (Highsmith 2002), [55]. Williams and Cockburn [18] state that agile development is ''about feedback and change'', that agile methodologies are developed to ''embrace, rather than reject, higher rates of change''.

Agility is the ability to sense and response to business prospects in order to stay inventive and aggressive in an unstable and rapidly shifting business environment (Highsmith, 2002), [55]. The agile approach to development is about agility of the development process, development teams and their environment (Boehm & Turner, 2004), [39]. This approach incorporates shared ideals of various stakeholders, and a philosophy of regular providing the customers with product features in short time-frames (Southwell, 2002), [45]. This frequent and regular feature delivery is achieved by team based approach (Coram & Bohner, 2005), [47].

Agile teams consist of multi-skilled individuals (Fowler, 2002), [16]. The development teams also have on-site customers with substantial domain knowledge to help them better understand the requirements (Abrahamsson, Solo, Ronkainen, & Warsta, 2002), [37]. Multiple short development cycles also enable teams to accommodate request for change and provide the opportunity to discover emerging requirements (Highsmith, 2002 ), [55]. The agile approach promotes micro-project plans to help determine more accurate scheduling delivery commitments (Smits, 2006), [48].

M Lindvall, V Basili, B Boehm, P Costa, (2002), [17] summarize the working definition of agile methodologies as a group of software development processes that must be iterative (take several cycles to complete), incremental (not deliver the entire product at once), self-organizing (teams determine the best way to handle work), and emergent (processes, principles, and work structures are recognized during the project rather than predetermined). In the paper by (Abrahamsson, Warsta, Siponen & Ronkainen, 2003), in general, characterized agile software development by the following attributes: incremental, cooperative, straightforward, and adaptive [24]. Boehm, B., & Turner, R. (2005), generalize agile methods are lightweight processes that employ short iterative cycles, actively involve users to establish, prioritize, and verify requirements, and rely on a team's tacit knowledge as opposed to documentation [30].

## 2. Agile Methods

For over a decade now, there has been an ever increasing variety of agile methods available includes a number of specific techniques and practices of software development. Agile methods are a subset of "iterative and evolutionary methods" [83, 84] and are "based on iterative enhancement" [85] and "opportunistic development processes" [86]. Most of agile development methods promote development, teamwork, collaboration, and process adaptability throughout the life-cycle of the project [4].

The major methods include eXtreme Programming (Beck, 1999), [82], Scrum (K. Schwaber & Beedle, 2002), [53], Dynamic Systems Development Method (Stapleton, 1997), Adaptive Software Development (Highsmith, 2000), Crystal (Cockburn, 2002), and Feature-Driven Development (Palmer & Felsing, 2002). [58], [59], [60], [61]. Figure 1 shows an agile software development methodology process flow (Scrum).

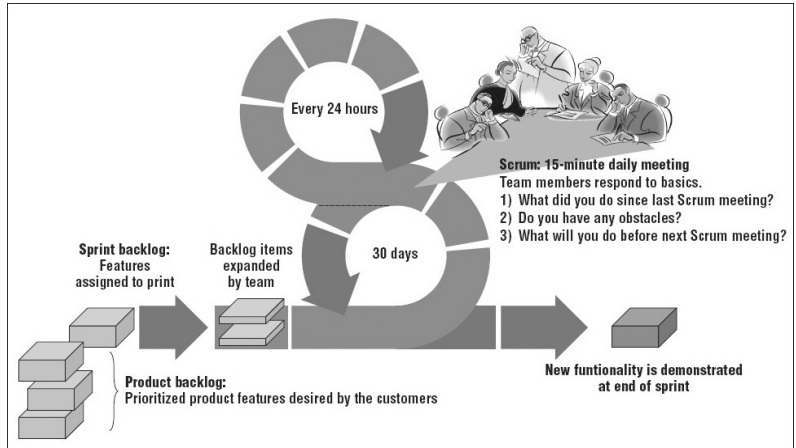

Figure 1: An example of agile software development methodology: Scrum (Source: [53])

The Agile Manifesto articulates the common principles and beliefs underlying these methods (Cockburn, 2002), [16]. Among the first and perhaps best known agile methods are Scrum and XP (Salo, & Abrahamsson, 2008), [49]. See Figure 2 shows the current rate of Agile methodologies used. Scrum is aimed at providing an agile approach for managing software projects while increasing the probability of successful development of software, whereas XP focuses more on the project level activities of implementing software. Both approaches, however, embody the central principles of agile software development [31].

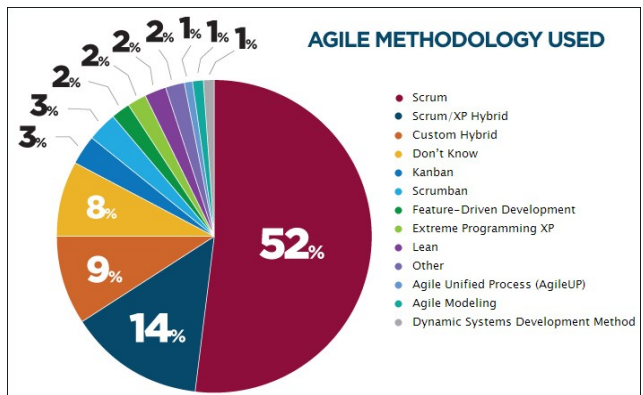

Figure 2: State of Agile Survey Results 2011 by VersionOne Inc.
(Source: http://www.versionone.com [10] )

*Title*

Agile software development processes -- such as the Rational Unified Process (RUP), Extreme Programming (XP), Agile Unified Process (AUP), Scrum, Open Unified Process (OpenUP), and even Team Software Process (TSP) -- are all iterative and incremental (evolutionary) in nature [63]. Some these modern approaches, in particular XP and Scrum, are agile in nature. The agile methods are focused on different aspects of the software development life cycle. Some focus on the practices (extreme programming, pragmatic programming, agile modeling), while others focus on managing the software projects (the scrum approach) [12].

## 3. Comparison Agile software development methodologies over traditional SDMs

There are many different characteristics between ASDMs and TSDMs. Boehm [16], for example, reports nine agile and heavyweight discriminators. He believes the primary objective of ASDMs is on rapid value whereas the primary objective of TSDMs is on high assurance.

Study performed S. Nerur, R. Mahapatra, G. Mangalaraj [22] state a comparison of traditional and agile development, they report seven issues to differentiate traditional and agile development. Their fundamental assumption of traditional development: "system are fully specifiable, predictable, and are built through meticulous and extensive planning" , whereas agile development: "high-quality adaptive software is developed by small teams using the principles of continuous design improvement and testing based on rapid feedback and change".

T. Dyba, & T. Dingsoyr, [74] summarize the differences between Agile development and traditional development basis on the of an unpredictable world, as well as emphasizing the value competent people and their relationships bring to software development. Agile methods address the challenge of an unpredictable world, emphasizing the value competent people and their relationships bring to software development [74].

Different researchers compare traditional and agile approaches, in their different perspectives, are summarized in Table 2 (All sources from additional information).

| Issues | Traditional Approach | Agile Approach |
|---|---|---|
| **Development life cycle** (Charvat, 2003); (Nerur, Mahapatra, & Mangalaraj,2005),[34],[22] | Linear; Life-cycle model (waterfall, spiral or some variation) | Iterative; The evolutionary-delivery model |
| **Style of development** (Leffingwell, 2007), [50] | Anticipatory | Adaptive |
| **Requirements** (Boehm, 2002); (Boehm and Turner, 2004), [16], [39] | Knowable early, largely stable; Clearly defined and documented | Emergent, rapid change, unknown – Discovered during the project |
| **Architecture** (Boehm, 2002); (Wysocki, 2009, 2011), [16], [56] | Heavyweight architecture for current and future requirements | YAGNI precept ("You aren't going to need it") |
| **Management** (Boehm, & Turner, 2005), (Vinekar, Slinkman,& Nerur, 2006),[30], [51] | Process-centric; Command and control | People-centric; Leadership and collaboration |
| **Documentation** (Boehm and Turner, 2005) ,[30] | Heavy / detailed Explicit knowledge | Light (replaced by face to face communication) Tacit knowledge |
| **Goal** (Dybå & Dingsøyr, 2009), [74] | Predictability and optimization | Exploration or adaptation |
| **Change** (Boehm and Turner, 2003), [19] | Tend to be change averse | Embrace change |
| **Team members** (Boehm, 2002) , (Sherehiy, Karwowski, & Layer, 2007), [16], [41] | Distributed teams of specialists; Plan-oriented, adequate skills access to external knowledge | Agile, knowledgeable, collocated and collaborative; Co-location of generalist senior technical staff; |
| **Team organization** (Leffingwell, 2007), [52] | Pre-structured teams | Self-organizing teams |
| **Client Involvement** (Highsmith & Cockburn, 2001), [21] | Low involvement; Passive | Client onsite and considered as a team member; Active/proactive |
| **Organization culture** (Highsmith, 2002) , (Nerur, Mahapatra, Mangalaraj, 2005), [55], [22] | Command and Control Culture | Leadership and Collaboration Culture |
| **Software development process** (Salo, & Abrahamsson, 2007), [42] | Universal approach and solution to provide predictability and high assurance | Flexible approach adapted with collective understanding of contextual needs to provide faster development |
| **Measure of success** (Highsmith, 2010), [1] | Conformance to plan | Business value delivered |

Table 2: Traditional and agile perspectives on software development (Sources: from literature review).

*Title*

## 3.2 Major Agile benefits in comparison to the traditional approach

In this section, we presenting list and explain some of agile benefits in comparison to the traditional approach which significantly improves software development in many ways. We try to provide an in-depth understanding (in some cases with figures), of these merit issues:

### 3.2.1 Evolutionary approach

Agile software development is a highly collaborative and evolutionary approach [101]. Agile methods become more popular in the software development industry. In their different research papers, (Boehm, & Turner, 2005; Larman, & Basili, 2003; Greer, & Ruhe, 2004; Dybå, & Dingsøyr, 2008; Paetsch, Eberlein, 2003; Abrahamsson, Warsta, 2003; Dagnino, 2002 ), they believe, Agile methods are iterative, evolutionary, and incremental -delivery model of software development [30], [79], [29] ,[20] ,[80] ,[24] ,[81].

Entire application is distributed in incremental units called as iteration. Development time of each iteration is small (couple of weeks), fixed and strictly adhered to. Each iteration is a mini increment of the functionality and is build on top of previous iteration. Agile software development of short iterative cycles offers an opportunity for rapid, visible and motivating software process improvement [75].Traditional approaches to the data-oriented aspects of software development; however, tend to be serial, not evolutionary and certainly not agile, in nature.

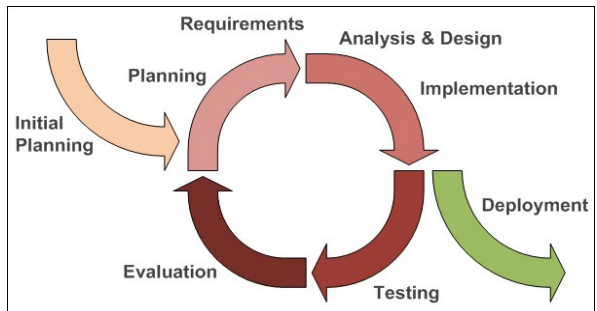

Figure 3: Iterative and incremental agile development process
(source: agile-development-tools.com).

### 3.2.2 Lightweight Methods

Boehm, B., & Turner, R. (2005), generalize agile methods are lightweight processes that employ short iterative cycles, actively involve users to establish, prioritize, and verify requirements, and rely on a team's tacit knowledge as opposed to documentation [30]. G Perera, & MSD Fernando (2007), also describe Agile practice is a customer oriented, light-weight software development paradigm, best suited for small size development teams in projects under vague and changing requirements [65]. A number of agile software development methods such as extreme programming (XP), feature-driven development, crystal clear method, scrum, dynamic systems development, and adaptive software development, fall into this category [22]. Traditional Software

Development Methods (TSDMs) including waterfall and spiral models are often called heavyweight development methods [26]. These methods involves extensive planning, predefine process phases, heavy documentation and long term design process. Lightweight methodologies put extreme emphasis on delivering working code or product while downplayning the importance of formal process and comprehensive documentation [23].

### 3.2.3 Rapid delivery of software products

Agile development methodologies emphasize rapid delivery of software products to the clients. According to (Boehm & Turner, 2005), Fast cycles, frequent delivery: Scheduling many releases with short time spans between them forces implementation of only the highest priority functions, delivers value to the customer quickly, and speeds requirements emergence [30]. ASD methods are iterative and incremental development [4], and each successful completion of development iteration, it delivers software product increment to client, thus Agile software development is satisfying the customer through early and continuous delivery of the valuable software [66]. Traditional, lifecycle based software development delivers the software only after entire completion of development process and before that clients have no clear idea and view of software to be developed.

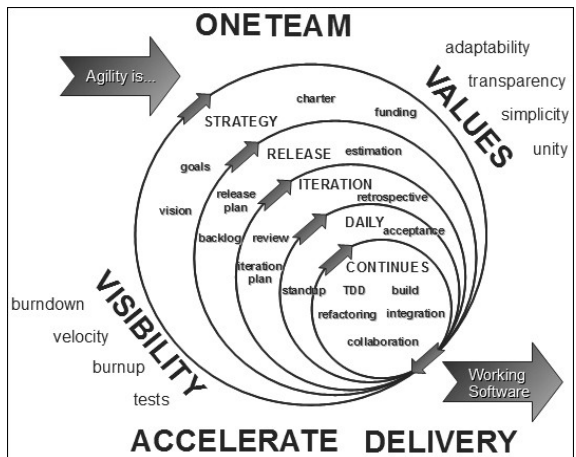

Figure 4: Iterative process and incremental delivery software products (source: [4]).

### 3.2.4 Highly tolerant of change requirements

The main difference between heavyweight and agile methodologies is the acceptance of change. It is the ability to respond to change that often determines the success or failure of a software project [18]. Heavyweight methods freeze product functionality and disallow change. Agile systems development methods emerged as a response to the inability of previous plan-driven approaches to handle rapidly changing environments (Highsmith, 2002). As second principle of Agile Manifesto [1] - "welcome changing requirements, even late in development", all agile method(s) is well organized,

*Title*

accommodate to change requirements. According to B. Boehm, (2002), organizations "are complex adaptive systems in which requirements are emergent rather than pre-specifiable" and agile approaches "are most applicable to turbulent, high- change environments" [16]. Agile software development promotes adaptive planning, evolutionary development and delivery, and encourages rapid and flexible response to change [4]. See Figure 5.

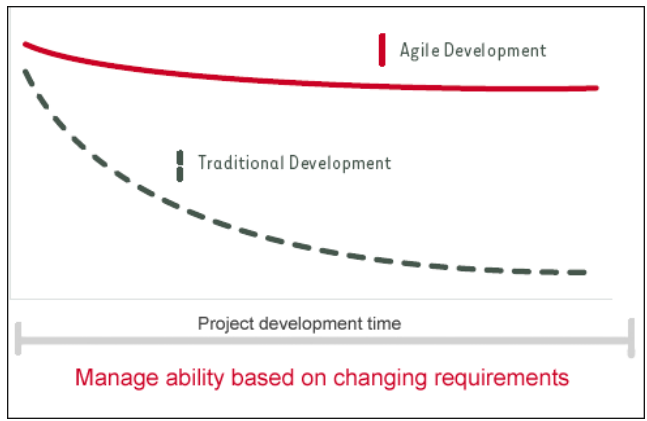

Figure 5: Agile vs. traditional requirements change management (Source: www.versionone.com)

Agile development inherently welcomes requirement changes as well as inclusion or exclusion of features throughout the development lifecycle. It is possible to accept requirement changes while in development phases because of iterative developments involve with agile development approach. As a result of this iterative planning and feedback loop, teams are able to continuously align the delivered software with desired business needs, easily adapting to changing requirements throughout the process. See Figure 6.

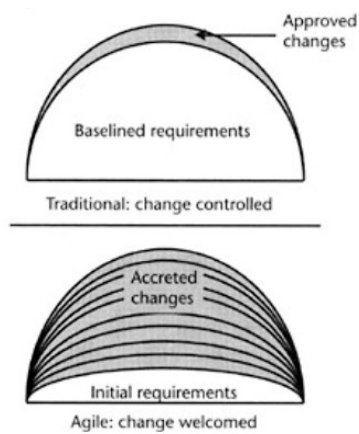

Figure 6: Agile vs. traditional requirements change management (source: [57]).

In contrast, agile development framework allows both customers and developers to change the requirements throughout the project, but only the customers have the authority to approve, disapprove and prioritize the ever-changing requirements (Koch, 2005), [57]. In traditional SDMs it increases complexity for accepting changing requirements while developing, and also increases development and delivery time, as well as cost to deliver software product.

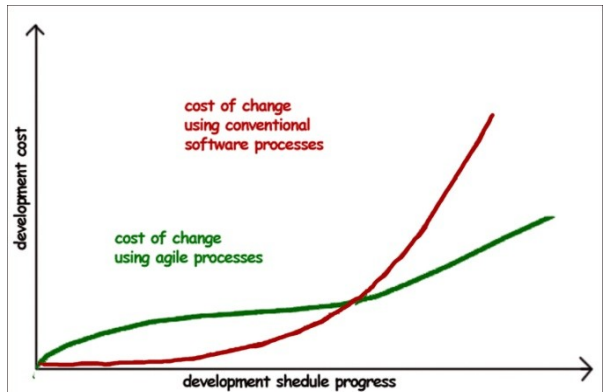
Figure 7: Cost of change for agile and conventional development process.

### 3.2.5 Accept prioritizing requirements

In agile software development, requirements always provided by client and these requirement features are prioritized by client itself. Agile methods break development tasks into small increments with minimal planning and do not directly involve long-term planning. Iterations are short time phases that typically last from one to four weeks. Thus, top prioritized features can be delivered each of development iteration. Agile requirements prioritization techniques to support and deal with frequent changes in priority lists which have been identified as success issue to accommodate over changes [73]. In traditional development, software product with all features will be delivered at a time only after completion of software project.

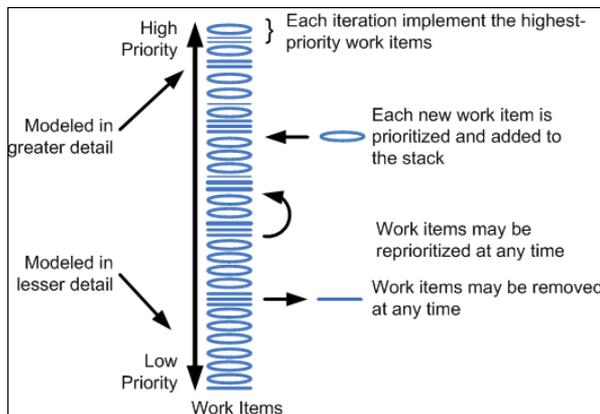
Figure 8: Agile approach prioritized requirements (Source: www.agilemodeling.com [6]).

### 3.2.6 Active customer involvement & feedback

Customers are actively involved, and get higher priority in agile approaches rather than any traditional approaches. There is face to face communication and continuous feedback from customer (product owner) always happen in agile approach.

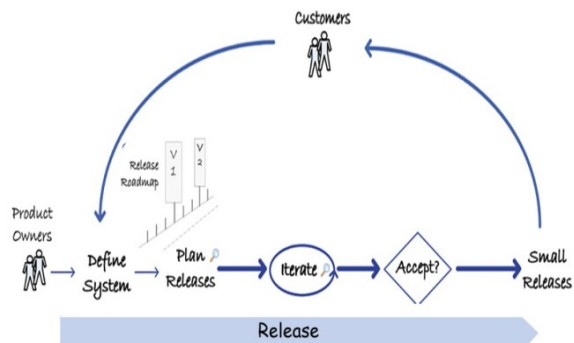

Figure 9: Active customer involvement in agile approach.

Customers appreciate active participation in projects as it allows them to control the project and development process is more visible to them, as well as, they are kept up to date [73]. This customer involvement mitigates one of the most consistent problems on software projects: "What they will accept at the end of the project differs from what they told us at the beginning". This interaction helps the customer to form a better vision of the emerging product. Along with the ability to visualize the functionality that is coming based on having seen what was built so far, the customers develop a better understanding of their own needs and the vocabulary to express it to the developers [9]. Agile projects require a meaningful client involvement in every part of the project to provide constant feedback in an open and honest way (Wysocki, 2009), [57]. This feedback is a key element of agile methodologies, which is why the customer must be committed, knowledgeable, collaborative, representative, and empowered to avoid risk of failure (Boehm, 2002), [16]. People are the primary drivers of agile projects and agile teams work best when people are physically close and document preparation and dissemination are largely replaced by face-to-face communication and collaboration (Cockburn & Highsmith, 2001), [21].

### 3.2.7 Reduce cost and time

The study reports conducted by B. Bahli and ESA Zeid [77] that the development team found using the waterfall model to be an ''unpleasant experience'', while XP (an agile method) was found to be ''beneficial and a good move from management''. The XP project was delivered a bit less late (50% time-overrun, versus 60% for the traditional), and at a significantly reduced cost overrun (25%, compared to 50% cost overrun for the traditional project). Agile development involves less cost of development as rework, management, documentation and other non-development work related cost is reduced.

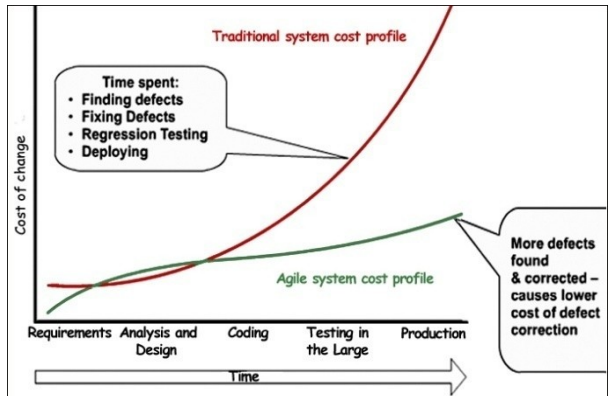

Figure 10: Cost for agile development process and conventional development.
(Source: http://www.thoughtworks.com).

**3.2.8 Short design phase involves early feedback from clients**

In traditional, lifecycle based developments usually follow Big Design Up Front and Big Requirements Up Front development techniques. With these approaches, comprehensive requirements document and design document are developed early in the project lifecycle which is used to guide the design and implementation efforts. It is typically months, if not years, before stakeholders are shown working software which implements their requirements and design. In terms of the traditional project phases (requirements, analysis, architecture, design) these take sixty percentage development time of project and still then there is no working software is ready for the client feedback.

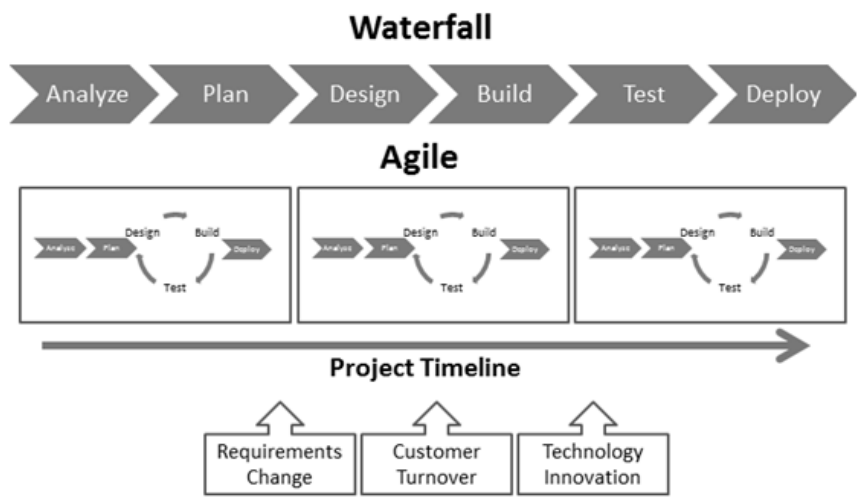

Figure 11: design phase composition between waterfall and agile development.

*Title*

According to (Boehm & Turner, 2005), agile approach design is simple which involves Designing for the battle, not the war. The motto is YAGNI (You Aren't Going to Need It). The antimotto is BDUF (Big Design Up Front). Strip designs down to cover just what you're developing. Since change is inevitable, planning for future functions is a waste of effort [30].Customer gets to know regular and frequent status of the application and delivery is defined by fixed timescale. So, customer is assured of receiving some functionality by a fixed time period. Due to the short development life cycle through an iterative and incremental process, the agile methods have been used widely in business sectors where requirements are relatively unstable [26].

### 3.2.9 Self organized team

Agile teams are self organizing and roles and relationships evolve as necessary to meet objectives (Leffingwell, 2007). Team composition in an agile project is usually cross-functional and self-organizing, without consideration for any existing corporate hierarchy or the corporate roles of team members [4]. Agile product development practices introduce changes in team culture in an attempt to bringing reciprocal effects of roalty and commitment to the team and projects (Sherehiy, Karwowski, & Layer, 2007). Team members normally take responsibility for tasks that deliver the functionality an iteration requires. They decide individually how to meet an iteration's requirements. Teams develop applications collaboratively and in cooperative environment. Agile alliance [5], claims that for a given problem size, "fewer people are needed if a lighter methodology is used, and more people are needed if a heavier methodology is used," and asserts that, "There is a limit to the size of problem that can be solved with a given number of people" [44].

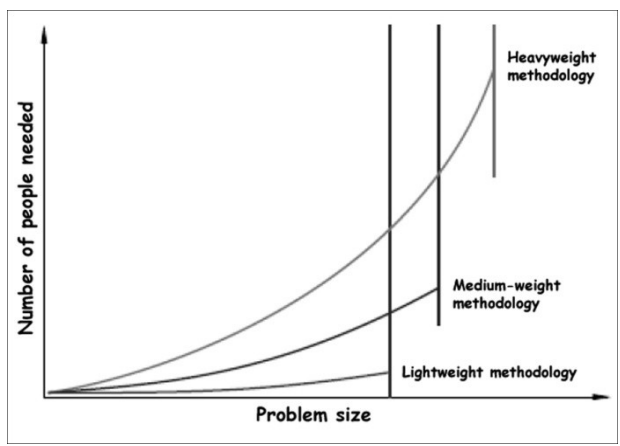

Figure 12: problem size; number of people needed (Source: Cockburn, 2007)

**3.2.10 Documentation**

Agile development improvement in productivity, reduction development cost and reduction in time-to-market (Reifer, 2002), [40]. Agile approaches, emphasis more is on developing the application only, and not on documentation. According to Wysocki, non-value-added work involves the consumption of resources (usually people and time) on activities that do not add business value to the final product or process [56]. Simple and minimal documents are used to exchange the views. Reducing intermediate artifacts that do not add value to the final deliverable means more resources can be devoted to the development of the product itself and it can be completed sooner.

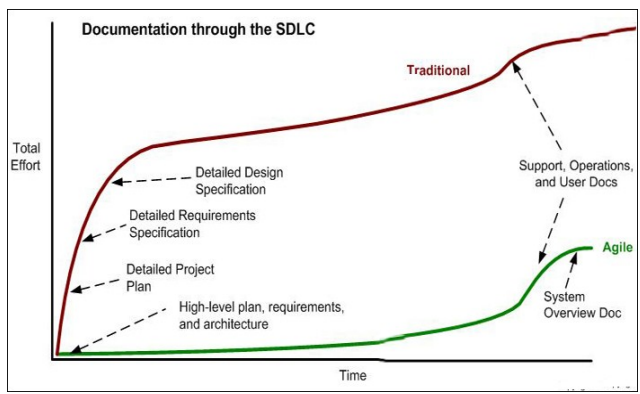

Figure 13: Agile vs. Traditional development documentation through the SDLC. (Source: www.agilemodeling.com [6]).

**3.2.11 Design simplicity**

According to (Boehm & Turner, 2005), agile approach design is simple which involves Designing for the battle, not the war. The motto is YAGNI (You Aren't Going to Need It). The anti-motto is BDUF (Big Design Up Front). Strip designs down to cover just what you're developing. Since change is inevitable, planning for future functions is a waste of effort [30]. In their research paper [46], ( K Molokken, & Ostvold, 2005 ), define agile method(s) as a flexible software development model(s), basis on evolutionary and incremental models; and also claim that, among the benefits of using these models are reduced software project overruns.

**3.2.12 Improves software quality**

Boehm, B., & Turner, R. (2004, May), Agile development methodologies (such as XP, Scrum, and ASD) promise higher customer satisfaction, lower defect rates, faster development times and a solution to rapidly changing requirements. Plan-driven approaches such as Cleanroom, the Personal Software Process, or methods based on the Capability Maturity Model promise predictability, stability, and high assurance [38].

*Title*

The regular and continuous interaction between the customer and the developers have as their primary objective assuring that the product as built does what the customer needs for it to do and assures the usability of the product as well. The strong technical focus results in much better testing on an Agile project than in most other methods [9]. According to Charvat, (2003), agile practices: iterative and adaptive life cycles have the advantage of a continual testing throughout the project, which has a positive impact on quality [43]. See Figure 2.

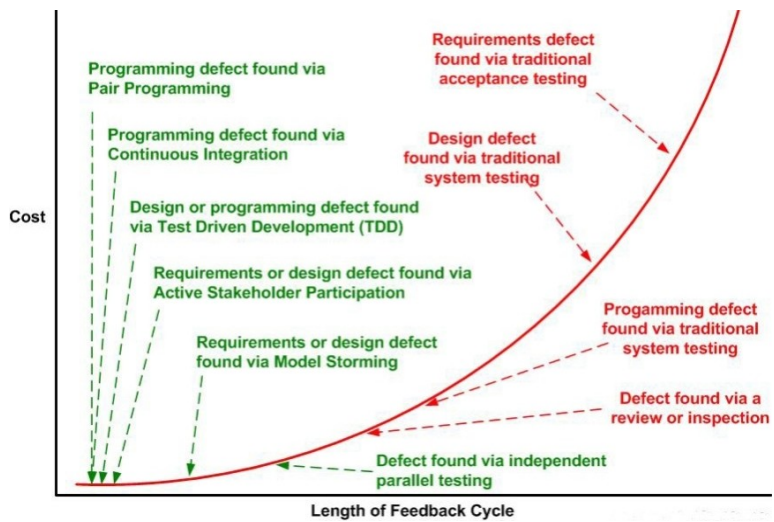

Figure 14: Comparison of Feedback cycles with traditional approaches. (Source: http://www.ambysoft.com)

Agile developers take responsibility for the quality of the code they write. In addition to producing cleaner code, it means that if there are testing specialists on the project, they will start their testing with better software, which always results in more effective testing and a better resulting product. In addition to, developers value the technical focus on testing and refactoring of agile methods increasing their motivation. There is also a perception of increased quality in software products and higher productivity when using some agile teams use practices like coding standards, peer reviews, and pair programming to assure that the code they produce is technically solid [73].

**3.2.13 Increase Business value, visibility, adaptability and reduce cost**

Agile software development accelerates the delivery of initial business value, and through a process of continuous planning and feedback, ensures that value continues to be maximized throughout the development process. ASD provides customer satisfaction through collaboration and frequent delivery of implemented features. By delivering working, tested, deployable software on an incremental basis, agile development delivers increased value, visibility and adaptability much earlier in the life cycle, significantly reducing project risk.

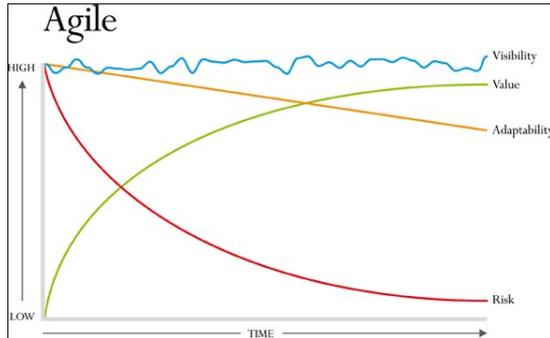

Figure 16: Agile development value proposition (Source: [10]).

### 3.2.14 Success possibility increased

According to various studies, almost 70% of all software projects fail. Materially fail to meet their objectives, in terms of cost, time, features, or all of the above. Traditional methods of managing software delivery have failed to deliver the predictability they promise. Agile practices benefit in terms of increased project success rate and user acceptance, better risk management, delivery of quality content on time and most important adjust to changing requirements [66]. See Figure 17.

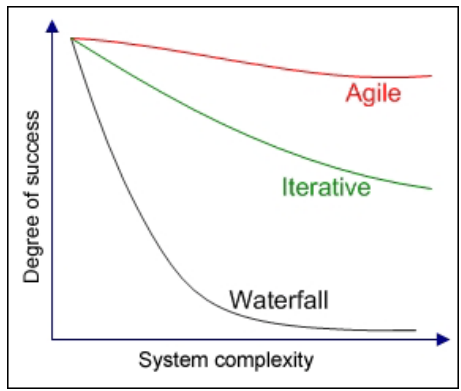

Figure 17: Agile development degree of success.

In a study by Boehm and Papaccio [72] discovered that a typical project experiences a 25% change in requirements, while yet another [Johnson] showed that 45% of features were never used. Agile approach aims to reduce waste and over-production by determining which parts are actually needed by the customer at each stage. In Agile approaches, delivering software on an incremental basis, customers give continuous feedback and agile team will always deliver products on time and on budget. As

*Title*

traditional project management isn't succeeding, more and more companies are turning to Agile development.

According to the Standish Group's, [11] famous CHAOS Report of 2000, 25% of all projects fail outright through eventual cancellation, with no useful software deployed. Sadly, this represents a big improvement over CHAOS reports from past years. Recently, they conduct a survey for Agile implementation success rate, see figure 19.

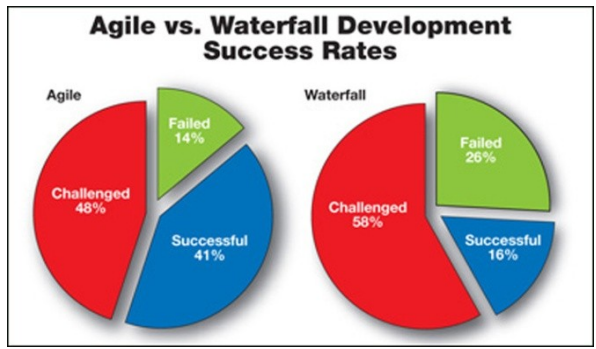

Figure 19: Agile implementation success rate by The Standish group, (Source: http://blog.standishgroup.com/) [11].

Survey result shows: most of the clients are asking for Agile implementation due to unprecedented benefits of Agile, over the other methodology, such as time to market, quality, defect rate, customer satisfaction, continuous end user feedback. This requires vendors to quickly turnaround and respond, to market demands, which eventually forces the organization to reevaluate the present onshore-offshore model.

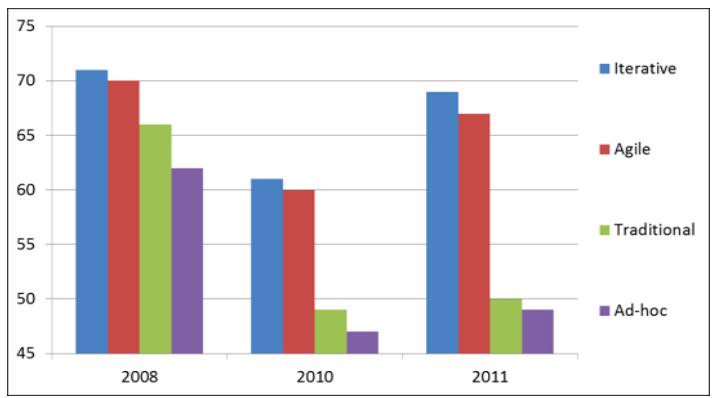

Figure 20: Agile projects success rate by Scott Ambler, (Source: www.ambysoft.com/surveys/)

Another survey conducted by Scott Ambler has consistently (2008, 2010 & 2011) shown that Agile and Iterative Projects have been more successful. Apart from the fact that Agile has been consistently been more successful compared to traditional approach.

## 4. Agile Adoption

Agile methods are highly being adopted because of expectations that these methods can bring development success (Esfahani, Yu, & Annosi, 2010). One of the main reasons for success with agile methods is that they are highly adaptive (Boehm & Turner, 2003), [38]. Figure 1 reveals the current levels of agile adoption. In this case, 71% of respondents indicated that they work in organizations that have succeeded at agile and an additional 15% work in organizations that have tried agile but have not yet succeed at it.

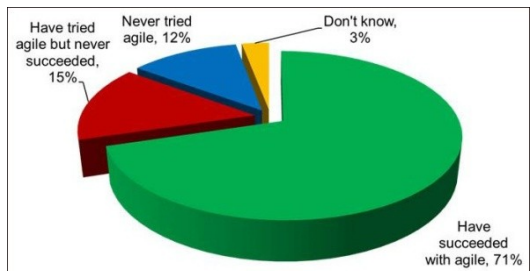

Figure 21: Agile adoption rates. (Source: http://www.ambysoft.com/surveys)

Salo, O., & Abrahamsson, P. (2008), argue that scientific publications and anecdotal evidence demonstrate that organizations worldwide are adopting agile software development methods at increasing speed [31]. In the study report, conducted by Forrester Research in 2011, agile development approaches adoption increases 35.4% to 38.6% whether as, traditional as well as, iterative approaches decreases. See figure 0.

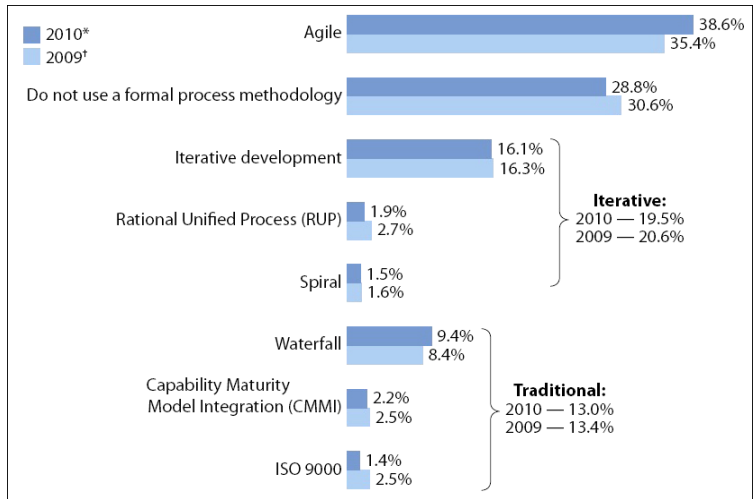

Figure 22: Forrester Research Agile Adoption rate rises. (Source: http://www.forrester.com [13])

*Title*

According to (West & Grant, 2010), "in the past few years, Agile processes have not only gained increasing adoption levels; they have also rapidly joined the mainstream of development approaches" [28]. Mary large companies including HP, IBM, Oracle, and Microsoft use Agile methodologies [76] — and more and more smaller organisations turn Agile each year. In their study (West & Grant, 2010), conducted by Forrester Research in 2009, agile software development processes were in use in 35% of organizations, and another 16% of organizations used an iterative development approach, while only 13% of organization use a Waterfall approach. However, nearly 31% did not use a formal development methodology [28].

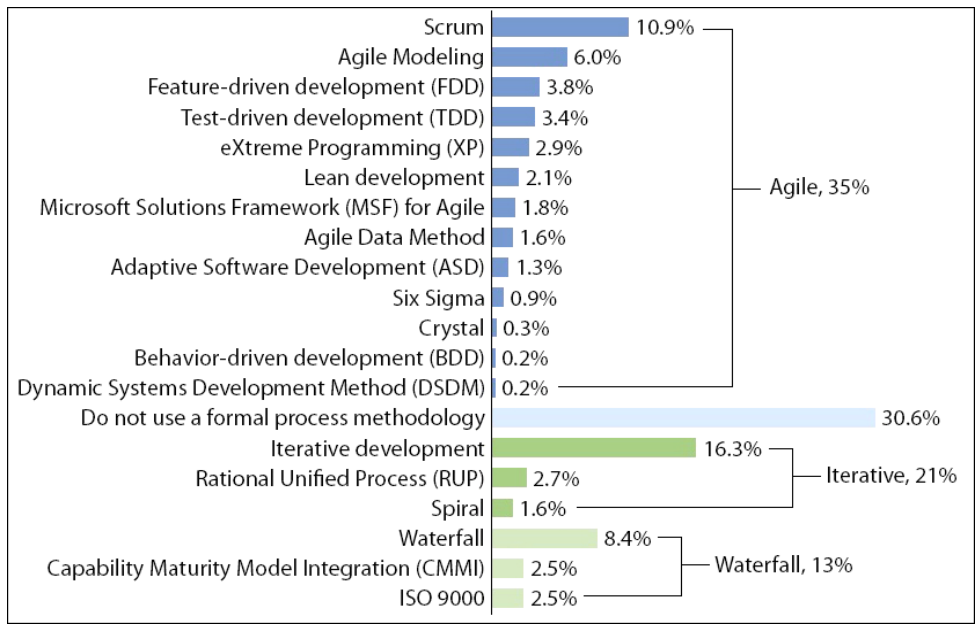

Figure 23: Agile adoption rates by Forrester Research in 2009 (Source: [28])

The main reasons behind for adopting Agile approaches rather than plan-driven approaches relate to: rapid changes; need for rapid results; emergent requirements (Boehm & Turner, 2003), [38]. According to Charvat, (2003), Leffingwell, (2007), & Perrin, (2008), Agile methodologies have numerous advantages including that they: adapt very well to change and dynamism; are people-oriented and value-driven, rather than process-oriented and plan-driven; mitigate risks by demonstrating values and functionalities up front in the development process; provide a faster time to market; improve productivity (by reducing the amount of documentation) and will fail early/quickly and painlessly, if a project is not doable [34], [33], [32].

A state of Agile survey 2011, conducted by versionone Inc. result shows: the top three reasons for adopting Agile to - accelerate time to market, increase productivity, and to more easily manage changing priorities.

*Author*

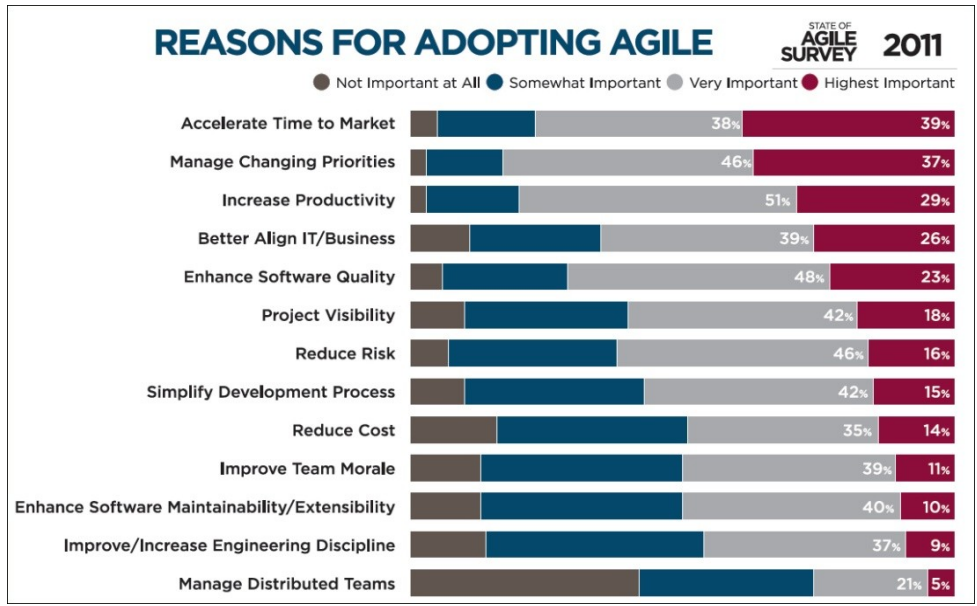

Figure 24: Reasons for adopting Agile from "A state of Agile survey 2011" (Source: www.versionone.com)

Prior to adoption, respondents said productivity and time to market ranked as their top reasons to adopt agile. But experienced agile users said actual benefits were primarily project visibility (77%) and the ability to manage changing priorities (84%).

# 5. Conclusion

Agile software development methodologies are evolutionary and incremental models have become increasingly popular in software development industry. Through, in many organizations, agile system development methods at adoption stage, agile methods might start to become well-established processes of these small, mid-level, even large organizations. There is increasing need to have a deeper understanding of agile methods in use in software development industry; as well as, have a better understanding – the benefits of agile approach as for accepting agile methods into their development style and for cope-up with their dynamic business needs.

In this paper, we present main issues of agile numerous benefits in comparison to the traditional approach which significantly improves software development process in many ways. We also provide with this paper, the current adoption state of Agile software development with different current survey results with graphs. The purpose of this paper is to provide an in-depth understanding- the benefits of agile development approach into the software development industry, as well as provide a comparison study report of ASDM over TSDM.

*Title*

*Title*

*Title*

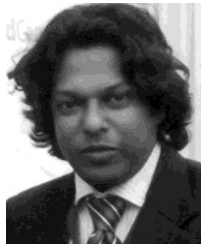

**A B M Moniruzzaman** Received his B.Sc (Hon's) degree in Computing and Information System (CIS) from London Metropolitan University, London, UK and M.Sc degree in Computer Science and Engineering (CSE) from Daffodil International University, Dhaka, Bangladesh in 2005 and 2013, respectively. Currently he is working on research on Cloud Computing and Big Data Analytics as a research associate at RCST (Research Center for Science and Technology) at Daffodil International University (DIU), Dhaka, Bangladesh. Besides, his voluntarily works as reviewer of few international journals including IEEE, Elsevier and IGI-Global. He is a student member of IEEE. His research interests include Cloud Computing, Cloud Applications, Open-source Cloud, Cloud Management Platforms, Building Private and Hybrid Cloud with FOSS software, Big Data Management, Agile Software Development, Hadoop, MapReduce, Parallel and Distributed Computing, Clustering.

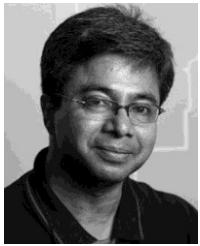

**Prof. Dr. Syed Akhter Hossain** Prof. Dr. Syed Akhter Hossain is Post Doctoral Fellow, Informatics and Systems Engineering, LIESP Laboratory, Universite Lyon2, Lyon, France. He received PhD in Computer Science and Engineering from University of Dhaka, Bangladesh and MSc degree in Applied Physics and Electronics, First Class (First), and BSc (Hons) in Applied Physics and Electronics, First Class (First), Gold Medalist from Rajshahi University, Rajshahi, Bangladesh. Currently he is working as Professor and Head, Department of Computer Science and Engineering, Daffodil International University, Dhaka, Bangladesh. Besides, he received best professor award from Singapore and has got more than 60 international publications including journals and proceedings and 3 book chapters with IGI Global and John Wiley. He is a member of ACM, and member of IEEE. His research areas includes simulation and modeling distributed system design and implementation, signal and image processing, internet and web engineering, network planning and management, database and data warehouse modeling.